\documentclass[11pt]{article}

\usepackage{jcappub,natbib,float,caption,bm}

\usepackage{amssymb}
\usepackage{amsmath}
\usepackage{hyperref}
\usepackage{longtable}
\usepackage{graphicx}
\usepackage{graphics}
\usepackage{dcolumn}
\usepackage{color}
\usepackage{rotate}
\usepackage{fancyhdr} 
\usepackage{comment}
\usepackage{empheq}

\def\be{\begin{equation}}
\def\ee{\end{equation}}
\def\bea{\begin{eqnarray}}
\def\eea{\end{eqnarray}}

\def\ba#1\ea{\begin{align}#1\end{align}}
\def\bcom#1\ecom{}
\newcommand{\vs}{\nonumber\\}

\def\cH{\mathcal{H}}
\def\E{\mathcal{E}}
\def\v#1{\bm{#1}}
\def\<{\left\langle}
\def\>{\right\rangle}
\def\comment#1{}

\DeclareMathOperator{\diag}{diag}

\def\vx{\v{x}}
\def\d{\delta}

\renewcommand{\refeq}[1]{Eq.~(\ref{eq:#1})}          
\newcommand{\refeqs}[2]{Eqs.~(\ref{eq:#1})--(\ref{eq:#2})}          
          
\newcommand{\refsec}[1]{Sec.~\ref{sec:#1}}
\newcommand{\refapp}[1]{App.~\ref{app:#1}}

\newcommand{\tn}{\textnormal}
\newcommand{\af}{{a_F}}

\newcommand{\mn}{{\mu\nu}}

\newcommand{\pd}{\partial}

\newcommand{\diff}{{\rm d}}

\newcommand{\bba}{\bar{a}}

\newcommand{\vt}{\Theta}

\newcommand{\oo}{{[1]}}

\newcommand{\dl}{\delta_L^\oo}

\newcommand{\daf}{\dot{a}_F}

\def\rhob{\bar\rho}

\title{Large-Scale Tides in General Relativity}

\author[a]{Hiu Yan Ip,}

\author[a]{Fabian Schmidt}

\affiliation[a]{Max-Planck-Institut f\"ur Astrophysik, Karl-Schwarzschild-Str. 1, 85741 Garching, Germany}

\abstract{Density perturbations in cosmology, i.e. spherically symmetric adiabatic perturbations of 
a Friedmann-Lema\^itre-Robertson-Walker (FLRW) spacetime, are locally exactly equivalent
to a different FLRW solution, as long as their wavelength is much larger
than the sound horizon of all fluid components.  
This fact is known as the ``separate universe'' paradigm.  However, no
such relation is known for anisotropic adiabatic perturbations,
which correspond to an FLRW spacetime with large-scale tidal fields.  
Here, we provide a closed, fully relativistic set of evolutionary equations 
for the nonlinear evolution of such modes, based on the \emph{conformal
Fermi (CFC)} frame.  We show explicitly that the tidal effects are
encoded by the Weyl tensor, and are hence entirely different from an
anisotropic Bianchi~I spacetime, where the anisotropy is sourced
by the Ricci tensor.  
In order to close the system, certain higher derivative
terms have to be dropped.  We show that this approximation is equivalent
to the \emph{local tidal approximation} of Hui and Bertschinger \cite{Hui:1995bw}.  We also
show that this very simple set of equations matches the exact evolution of 
the density field at second order, but fails at third and higher
order. This provides a useful, easy-to-use framework for computing
the fully relativistic growth of structure at second order.}

\begin{document}

\maketitle
\flushbottom

\section{Introduction}
\label{sec:intro}

Cosmological observations ranging from the cosmic microwave background (CMB)
to the clustering of galaxies strongly suggest that our universe is well described
by a Friedmann-Lema\^itre-Robertson-Walker (FLRW) spacetime on large
scales, with metric
\be
ds^2 = a^2(\tau) [-d\tau^2 + \gamma_{ij} dx^i dx^j]\,.
\label{eq:FLRW}
\ee
Here $a(\tau)$ is the scale factor, $\tau$ is conformal time  (with $d\tau = dt/a$), and $\gamma_{ij}$
is the metric of a maximally symmetric three-dimensional space with either
positive, negative, or vanishing curvature.  
However, observations also prove that there are perturbations to the
spacetime \refeq{FLRW} whose treatment is essential in order to correctly 
interpret cosmological observations.  

Interestingly, a special class of perturbations of \refeq{FLRW}, 
namely spherically symmetric adiabatic perturbations, are locally exactly equivalent
to a different FLRW solution $a \to a_F$, where $a_F$ obeys the Friedmann
equations, as long as their wavelength is much larger
than the sound horizon of all fluid components.  This is commonly known
as the ``separate universe'' picture \cite{lemaitre:1933,barrow/saich:1993,cole:1997,mcdonald:2003,sirko:2005,gnedin/kravtsov/rudd:2011,baldauf/etal:2011,sherwin/zaldarriaga,li/hu/takada:2014,wagner/etal:2014,baldauf/etal:2015}, and holds
at fully nonlinear order \cite{Dai:2015jaa}.  That is, an observer within
such a ``perturbed'' spacetime cannot distinguish the spacetime from exact FLRW by any local measurements, where local means 
on spatial scales much smaller than the wavelength of the perturbation.  Note that
this holds for all time, i.e. the observer could keep making spatially
local measurements for several Hubble times and would continue to find
agreement with FLRW.

However, no such relation to an exact solution of Einstein's equations 
is known for \emph{anisotropic} adiabatic perturbations around FLRW.  
While the spherically symmetric perturbations mentioned above can be
equivalently seen as density or curvature perturbations, the anisotropic
case is equivalent to large-scale tidal fields.  It is sometimes argued
that the exact solution corresponding to this case is a Bianchi~I spacetime,
\be
ds^2 = a^2(\tau) \left[-d\tau^2 + \left[\delta_{ij} + H_{ij}(\tau)\right]dx^i dx^j \right]\,,
\label{eq:BianchiI}
\ee
where $\delta_{ij}+ H_{ij}(\tau)$ is a trace-free symmetric positive-definite matrix.  Indeed, the motion of a non-relativistic test particle in such a spacetime is equivalent to that in an FLRW metric perturbed, in conformal-Newtonian gauge, by a tidal potential perturbation that can locally be written as (e.g., Sec.~7 of \cite{CFCpaper})
\be
\Psi \equiv -\frac12 a^{-2} \delta g_{00} = -\frac12 a^2 \left(\ddot H_{ij} + 2 H \dot H_{ij}\right) x^i x^j \,,
\label{eq:PsiBianchi}
\ee
where dots denote derivatives with respect to time $t$.  
However, as we will show below, this is not the correct physical picture.  
In a dust-filled universe (and indeed any universe dominated by an 
ideal fluid), the Bianchi~I solution \refeq{BianchiI} leads to a
rapidly decaying anisotropy $H_{ij} \propto a H$ (see \refsec{BianchiI}).  
Any other behavior would require significant
anisotropic stress.  The actual tidal fields in our universe on the other hand
grow in conjunction with the perturbations in the matter density.  
It follows that the tidal fields in our universe are not sourced
by any component of the Ricci tensor, but are instead encoded by the Weyl
tensor.  We show this by providing a closed, fully relativistic set of evolutionary equations 
involving the matter density, velocity shear, local scale factor $a_F$, 
and local tidal field $\E^F_{ij}$. $\E^F_{ij}$ is a specific component of the Weyl tensor. It can  be interpreted as the electric Weyl tensor evaluated on the geodesic or simply the "local tidal field" (motivated by the expression for the $00-$component of the metric perturbation in our framework, shown later). Specifically, these quantities
are defined in the \emph{conformal Fermi (CFC)} frame \cite{CFCorig,CFCpaper}, which ensures that
each one of them is a local observable from the point of view of an observer
comoving with the matter fluid.

Beyond elucidating the physical interpretation of tidal fields in the relativistic context, the result is also useful for estimating post-Newtonian corrections appearing in nonlinear cosmological perturbation theory.  The separate
universe picture proves that all local gravitational effects of isotropic 
adiabatic metric perturbations (at all orders in perturbation theory) 
are captured by the spatial curvature in the comoving frame \cite{Dai:2015jaa}.  
Our results prove that the corresponding quantity for anisotropic perturbations
is the electric part of the Weyl tensor.  Moreover, written in terms of
locally observable quantities, the evolutionary equations only contain terms
that are familiar from the subhorizon, Newtonian calculation (although,
of course, they do contain post-Newtonian terms once expressed in a specific
gauge such as Poisson gauge).  
However, in order to a obtain closed set of equations, certain terms need to
be dropped.  This \emph{local tidal approximation (LTA)} \cite{Hui:1995bw}
only recovers the correct physical evolution of the tidal field at linear order, and the density field at second order.  Hence, 
anisotropic (tidal) adiabatic perturbations are significantly more complex than
isotropic (density) perturbations.

The outline of this paper is as follows.  
In Section \ref{sec:CFC},  we briefly recap the idea of Conformal Fermi Coordinates (CFC). 
In Section \ref{sec:eqns}, we derive a closed system of equations for the  nonlinear evolution of the density and tidal fields in the CFC frame from the corresponding covariant system of equations.
In Section \ref{sec:sol}, we present the perturbative solutions for the quantities in the system up to second order and compare them to the known solutions in standard perturbation theory. We also relate the system to the collapse of a homogeneous irrotational ellipsoid and point out that, when restricting to the leading order perturbation in the CFC metric, our system matches the Local Tidal Approximation introduced by Hui and Bertschinger \cite{Hui:1995bw} for ellipsoidal collapse.
In Section \ref{sec:bispectrum}, we consider the rest-frame matter three-point function as an application of our results.
In Section \ref{sec:concl}, we summarize our findings and discuss their applications.

\section{Why a Bianchi~I spacetime does not describe large-scale perturbations in our Universe}
\label{sec:BianchiI}

We begin from the Bianchi~I metric \refeq{BianchiI}, and rotate the 
spatial coordinates to the frame where $H_{ij}$ is diagonal.  Then,
we can write
\be
\delta_{ij} + H_{ij}(t) = \exp \left(\begin{array}{ccc}
2\alpha(t) & 0 & 0\\
0 & 2\beta(t) & 0\\
0 & 0 & 2\xi(t)
\end{array}\right)\,,
\ee
where
\be
\alpha + \beta + \xi = 0
\ee
for all $t$.  In other words, we write \refeq{BianchiI} as
\be
ds^2 = -dt^2 + a^2(t) \left[ e^{2\alpha(t)} dx^2 + e^{2\beta(t)} dy^2 + e^{2\xi(t)} dz^2 \right]\,.
\label{eq:BianchiIb}
\ee
Jacobs~\cite{jacobs:68} derived the solution of the Einstein equations
for the ansatz \refeq{BianchiIb}, assuming a perfect-fluid stress tensor
(he also considered the case of a uniform magnetic field, which we do not discuss
here), and obtained [Eq.~(10) there]
\ba
\ddot \Pi + 3 H \dot\Pi =\:& 0\,,\quad \Pi := \alpha+\beta = -\xi\vs
\ddot \Sigma + 3 H \dot\Sigma =\:& 0\,,\quad \Sigma := \alpha-\beta\,,
\ea
where $H=\dot a/a$, while $\Pi$, $\Sigma$ are the 
anisotropy parameters.  The isotropic scale factor $a(t)$ satisfies the
standard Friedmann equation.  Thus, apart from an unobservable constant
mode, which can be removed by a trivial redefinition of spatial coordinates,
the anisotropy always decays as $\dot\Pi,\,\dot\Sigma \propto a^{-3}$.  
The tidal field experienced by a local comoving observer in this spacetime
[\refeq{PsiBianchi}] then scales as
\be
\partial_i\partial_j\Psi \propto a^2 \left[ \ddot\Pi + 2 H \dot\Pi\right] = - a^2 H \dot\Pi \propto aH(t)\,,
\ee
and analogously for $\Sigma$.  
That is, in a universe whose stress energy content is given by a perfect
fluid (more specifically, in the absence of significant anisotropic stress),
any initial tidal field described by a Bianchi~I spacetime decays rapidly, 
$\propto aH$.  In case of initial conditions from inflation, which are set
when a given mode exits the horizon, this contribution
disappears exponentially fast after horizon exit, as expected for a decaying mode.  
This is clearly very different from actual tidal fields, which do not
decay outside the horizon, and whose importance
in the growth of structure remains relevant up to late times.

\section{Recap of conformal Fermi coordinates}
\label{sec:CFC}

We begin by recapitulating the gist of the Conformal Fermi Coordinates (CFC), $\left\{t_F,x_F^i \right\}$, rigorously defined in \cite{CFCpaper}. Quantities defined with respect to CFC shall be denoted by a subscript $F$. Take an observer free-falling in some general spacetime.  His worldline is then a timelike geodesic in said spacetime.  For most applications of these coordinates, we take the spacetime to be perturbed Friedmann-Lema\^itre-Robertson-Walker (FLRW), although this is not a necessary assumption.  We construct a coordinate system centered on the observer, such that he always sees an unperturbed FLRW spacetime on the geodesic, with corrections going as the spatial distance from him squared, $\mathcal{O}\left[\left(x_F^i\right)^2\right]$.  Note that a power expansion in $x_F^i$ requires that $|x_F^i| $ be less than the typical scale of variation of $h_{F\mn}$, which we shall call $\Lambda^{-1}$. Since in a realistic cosmological setting, there exist metric perturbations on very small scales, we need to coarse-grain the metric (and the stress energy tensor) on a spatial scale $\Lambda^{-1}$. Then, the only contributing modes in the resulting metric perturbations have wavenumbers $k\lesssim \Lambda$.  CFC is then valid over a finite region \cite{CFCpaper}.  Since the absolute scale of the coarse-graining is not of relevance to the results of this paper, we will explicitly indicate the scale $\Lambda$.  

We can now proceed to construct the CFC frame for the coarse-grained metric. We start by taking the tangent $U^\mu = \diff x^\mu /\diff t_F$ to his worldine, i.e. the fluid 4-velocity, to be the time direction and  the hypersurface composed of all vectors orthogonal to his worldline to be the constant-time hypersurface, with the observer being at the spatial origin $x_F^i=0$. The orthonormal tetrad thus chosen at a point on his worldline is then parallel transported along the latter, such that these properties are preserved.  Then, given a scalar $a_F(x)$ that is positive in a neighborhood of the geodesics, the spatial coordinate for a given $t_F$ is defined as follows: $\left(t_F, x_F^i\right)$ is the point at which we arrive when starting from $P = \left(t_F, 0\right)$ (on the observer's worldline), we move along the spatial  geodesic of $\tilde{g}_\mn^F$ for a proper distance of $a_F(P) \sqrt{\delta_{ij} x_F^i x_F^j}$ in the direction defined w.r.t. the spatial components of the observer's tetrad.

For any given spacetime scalar $a_F(x)$\footnote{Note that $\af$ has to be defined not just on the geodesic, but in a neighbourhood around it in order for us to define CFC. \cite{CFCpaper}}, the metric can thus be made to take the form
\begin{equation}
g_\mn^F (x_F^\mu ) = a_F^2 (\tau_F) \left[\eta_\mn +h_\mn^F \left(t_F,x_F^i \right)\right],
\label{eq:CFCmetric}
\end{equation}
where 
\begin{align}\label{eq:CFCleading}
&h_{00}^F = -\tilde{R}_{0l0m}^F(\v{0}) x_F^l x_F^m\\
&h_{0i}^F = -\frac{2}{3}\tilde{R}_{0lim}^F(\v{0}) x_F^l x_F^m\\
&h_{ij}^F = -\frac{1}{3}\tilde{R}_{iljm}^F(\v{0}) x_F^l x_F^m.
\end{align}
Here, $\tilde{R}^F_{\mn \alpha\beta}(\v{0})$ is the Riemann tensor of the conformally-related metric,
\begin{equation}
\tilde{g}_\mn^F := a_F^{-2 } {g}_\mn^F \,,
\end{equation}
evaluated on the central geodesic.  

So far, we have not specified the local scale factor $a_F$.  As shown in
\cite{CFCpaper,Dai:2015jaa}, the natural, physically motivated choice is to define the
local Hubble rate along the geodesic through
\be
H_F(t_F) := a_F^{-1} \frac{\diff a_F}{\diff t_F} :=\frac{1}{3} \nabla_\mu U^\mu\,.
\ee
$a_F$ is then uniquely defined, up to an arbitrary overall multiplicative
constant, by integrating the Hubble rate along the observer's geodesic.  
Apart from reducing to $a_F=a$ for an unperturbed FLRW spacetime, this choice
ensures that $H_F$ as well as $h^F_\mn$ are strictly \emph{local observables}
from the point of view of the observer.\footnote{In fact, there is a residual gauge freedom to change $h^F_{ij}$ via a purely spatial gauge transformation, as discussed in \cite{CFCorig,CFCpaper}.  However, in this paper we will not deal with $h^F_{ij}$.}

 \section{Nonlinear evolution of density and tidal fields}
\label{sec:eqns}

In this section, we derive a closed system of evolutionary equations for 
the density $\rho$, velocity divergence $\Theta$, velocity shear $\sigma_{\mn}$,
and tidal field $E_{\mn}$.  Importantly, our relations will be derived
at fully nonlinear order and fully relativistically, without assuming $v^2 \ll c^2$ or subhorizon scales $k\gg \cH:= a H$ as usually done in large-scale structure studies.  Moreover,
by giving expressions in the CFC frame, our results correspond directly
to local observables from the point of view of a comoving observer.  

We will assume that the cosmological fluid is pressureless (CDM) and perfect. It follows  that there is no anisotropic stress and  the stress-energy tensor takes the form $T^\mn = \rho U^\mu U^\nu$, where $\rho$ is the proper energy density in the fluid rest frame.  
However, our results also hold in the presence of pressure, as long as pressure \emph{perturbations} can be neglected.  This holds trivially for a cosmological constant.  Moreover, it is valid as long as the long-wavelength perturbations considered are outside the sound horizon of all fluid components.  

Let us introduce the projection tensor $P_\mn (U) :=  g_\mn + U_\mu U_\nu$, which projects on the subspace orthogonal to the fluid 4-velocity (in CFC, $P_\mn$ will become trivial).  
The flow-orthogonal part of the velocity gradient can be decomposed as
\cite{1966ApJ...145..544H, 1971grc..conf..104E, 1971grc..conf....1E}
\be
P^\alpha{}_\mu  P^\beta{}_\nu \nabla_\alpha U_\beta 
= \frac{1}{3}\vt P_\mn +\hat{\sigma}_\mn\,,
\ee
where $\vt := P^\mn \nabla_\mu U_\nu$ is the expansion scalar. It describes the change in the volume of a sphere of test particles centered on the geodesic.  
The shear tensor $\hat{\sigma}_\mn$ is the traceless symmetric part of the velocity gradient tensor. It describes the rate of distortion of a sphere of test particles into an ellipsoid.  Here, we have neglected the vorticity $\omega_{\mu\nu}$, which is the antisymmetric part of the (flow-orthogonal) velocity gradient.  
Since, for a single barotropic fluid, vorticity is not generated, and any
initial vorticity decays, we set $\omega_{\mu\nu}$ to zero throughout.  Including this decaying mode is a trivial extension.

\subsection{Covariant equations}

Our goal is to generalize the separate universe picture, by finding a closed
system of equations for the evolution of a homogeneous ellipsoid.  
The Friedmann equations
governing the background evolution of an FLRW spacetime are a special
case of the Raychaudhuri equation, 
\begin{equation}\label{eq:gen1}
U^\alpha \nabla_\alpha\vt+\frac{1}{3}\vt^2+\hat{\sigma}_\mn\hat{\sigma}^\mn =  -4\pi G (\rho+3p)\,.
\end{equation}
While intrinsically a purely geometric relation derived from the Ricci identity,
we have used the Einstein equations to replace the Ricci scalar with the trace of the stress-energy tensor on the r.h.s..  The Ricci identity holds for any Levi-Civita connection $\Gamma^\alpha_{\mn}$ and is given by 
\begin{align}
\left[\nabla_\mu ,\nabla_\nu\right] U^\alpha = R^\alpha{}_{\beta\mn} U^\beta\,. 
\label{eq:Ricci}
\end{align}
Here, $\left[\nabla_\mu ,\nabla_\nu\right]$ is the commutator of the covariant derivatives and $R^\alpha{}_{\beta\mn} $ is the Riemann tensor of the metric $g_\mn$.  
It describes the difference between parallel-transporting $U^\alpha$ in the direction $\nabla_\mu$ then $\nabla_\nu$ and vice versa.  
Contracting \refeq{Ricci} with $P^\nu_{\  \alpha} U^\mu$ then yields \refeq{gen1}.  

We complement this with energy-momentum conservation, projected with the 4-velocity $U_\mu$,
\begin{equation}
\label{eq:gen3}
U_\mu\nabla_\nu T^\mn = 0,
\end{equation}
which for our CDM fluid  reduces to the evolutionary equation  for the rest-frame energy density, $\rho$.

In addition, we need an equation governing the evolution of the velocity
shear, which can again be derived from the Ricci idenity \refeq{Ricci}, 
namely by contracting with $U^\mu P^{\sigma\nu} P_{\rho \alpha}$ and then $P_{\mu\sigma}$.  This yields
\begin{equation}\label{eq:gen2}
U^\alpha \nabla_\alpha\hat{\sigma}_\mn = -\frac{2}{3} \vt \hat{\sigma}_\mn -\hat{\sigma}_{\mu\alpha}\hat{\sigma}^\alpha{}_\nu  +\frac{1}{3}P_\mn \hat{\sigma}_{\alpha \beta}\hat{\sigma}^{\alpha \beta}+ E_{\mu\nu} +\frac{1}{2}\hat{R}_\mn\,,
\end{equation}
where we have introduced
\be\label{eq:emn}
E_\mn (U) := C_{ \mu\alpha\nu \beta}U^\alpha U^\beta
\ee
and $C_{\alpha \nu\mu \beta}$ is the Weyl tensor \cite{Bertschinger:1993xt}, i.e. the traceless part of the Riemann tensor which describes the contributions of nonlocal sources to spacetime curvature.  As we will see, its \emph{electric part} $E_\mn$ can be understood as the invariant definition of the local tidal field.  Further, 
\be
\hat{R}_\mn := P^\alpha{}_\mu P^\beta{}_\nu R_{\alpha\beta} -\frac{1}{3} P_\mn P^{\alpha\beta} R_{\alpha\beta}
\ee
is the traceless part of the spatially-projected Ricci tensor.  \refeq{gen2}
shows that there are two sources of velocity shear: the electric
part of the Weyl tensor, and $\hat{R}_{\mn}$, which by virtue of the Einstein
equations is proportional to the trace-free part of the velocity-orthogonal
stress tensor $\hat{T}_{\mn}$.  In the absence of anisotropic
stress in the fluid rest frame, which is the case for baryons and 
cold dark matter, $\hat{R}_{\mn}$ vanishes.  Clearly, in the actual universe
\emph{velocity shear is sourced by the electric part of the Weyl tensor, which
is the relativistic generalization of the Newtonian tidal tensor}
$(\partial_i\partial_j - \d_{ij}/3 \nabla^2) \Psi$.  

Now, in order to obtain a closed system of evolutionary equations, we 
need an equation governing $E_\mn$.  This can be obtained by combining the
Bianchi identity and the Einstein equation,
\begin{equation}
\label{eq:gen4}
\nabla^\kappa C_{\mn \kappa\lambda}= 8\pi G \left(\nabla_{[\mu }T_{\nu ]\lambda}+\frac{1}{3}g_{\lambda[\mu }\nabla_{\nu ]}g_{\alpha\beta}T^{\alpha\beta}\right).
\end{equation}
This equation is the only one in the set that involves spatial---more precisely, fluid-orthogonal---derivatives 
acting on the Riemann tensor (via the Weyl tensor) and the density $\rho$.  
Note that fluid-orthogonal derivatives are simply spatial derivatives in the frame
comoving with the fluid.   
We are interested in describing the evolution of a long-wavelength perturbation.  In order to obtain a closed system of evolutionary equations, we thus \emph{discard fluid-orthogonal derivatives acting on the density and the Riemann tensor}  and neglect them in the following (of course, throughout we retain terms involving any derivatives acting on the velocity).  This is the key
approximation made in our derivation, and we will discuss its implications in detail below.

\subsection{Closed system in CFC frame}
\label{sec:eqns:CFC}

At this point, it is not obvious that the four relations Eqs.~(\ref{eq:gen1}), (\ref{eq:gen3}), (\ref{eq:gen2}), and (\ref{eq:gen4}) can be rewritten to form a closed local system.  However, this becomes clear once expressing all relations in the CFC frame.   
We denote $\dot{}= \frac{\diff}{\diff t_F}$ and $H_F := \frac{\daf}{\af}$.  We shall also adopt the convention where 3-indices are raised and lowered by the Kronecker delta, $\delta_{ij}$, whereas 4-indices are raised and lowered by $P_\mn$, which is the effective spatial metric for the observer in covariant form.  

Further, it is sufficient to evaluate all quantities on the geodesic $x_F^i=0$.  This is because we are free to choose a geodesic through any given point.  Thus, all terms that scale, after taking all spatial derivatives, as $\left(x_F^i\right)^n,\,\,n\geq 1$, vanish. This applies in particular to the peculiar velocity $v_F^i:= \frac{\diff x_F^i}{\diff t_F}=\mathcal{O}(x_F^i)$.  Then, without making any further approximations, the nonlinear tensor equations simplify greatly.  
On the geodesic, we have $U^\mu = a_F^{-1} (1,0,0,0)$, and the projection tensor simply becomes
\be
P^F_{\mn} = a_F^2\, \diag\left\{0,1,1,1\right\}\quad \tn{and}\quad 
P_F^{\mn} = \frac{1}{a_F^2}\diag\left\{0,1,1,1\right\}.
\ee
Firstly, \refeq{gen1} becomes 
\begin{equation}
\dot{H}_F+H_F^2 +\frac{1}{3 \af^2} \sigma_{Fij}\sigma_F^{ij} +\frac{4\pi G}{3} \rho_F=0\,,
\label{eq:gen1F}
\end{equation}
where
\begin{equation}
\sigma_{Fij}:= \frac{1}{\af} \left(P_F\right)^\mu{}_i \left(P_F\right)^\nu{}_j \hat{\sigma}_{F\mn} = \partial^F_{(i} v^F_{j)}\,,
\end{equation}
and $\partial^F_i \equiv \partial/\partial x_F^i$.  Note that we have pulled
out a factor of $a_F^{-1}$ in the definition of the velocity shear.  Further,
recall that in CFC, we define $\vt:= 3 H_F$ on the geodesic, such that the spatial velocity divergence, $\nabla_{k,F} v_F^k$, is absorbed into the definition of $\af$.  Thus, $\sigma_{Fij}$ is trace-free. 

Secondly, \refeq{gen3} becomes
\begin{equation}
\dot{\rho}_F  +3H_F \rho_F =0,
\label{eq:gen3F}
\end{equation}
which is unsurprisingly the familiar continuity equation.  
Thirdly, \refeq{gen2} becomes
\begin{align}
\dot{\sigma}_{Fij} + H_F \sigma_{Fij}+ \frac{1}{a_F}\left(\sigma_{Fil}\sigma_F^l{}_j -\frac{1}{3}\delta_{ij} \sigma_{Flm}\sigma_F^{lm}\right)+\frac{\E^F_{ij}}{a_F}=\:& 0\,, 
\label{eq:gen2F} \\
\mbox{where}
\quad \E^F_{ij} \equiv\:& a_F^{-2} C_{Fi0 j 0} \,.
\label{eq:EFdef}
\end{align}
Here, we have introduced $\E^F_{ij}$ as the relevant component of the Weyl
tensor in CFC frame, which is directly related to the perturbation $h^F_{00}$
of the 00-component of the CFC metric (see below).  
Further, we have used the fact that
\begin{align}
(P_F)^\mu{}_i\, (P_F)^\nu{}_j\,\hat{R}_{F\mn} = \hat{R}_{F ij} = R_{Fij} -\frac{1}{3} \delta_{ij} \delta^{kl}(R_F)_{kl} = 0\,.
\end{align}
This vanishes on the geodesic when we evaluate $R_{F\mn}$ using the trace-reversed Einstein equation because the CFC frame is defined to be the fluid rest frame, and we assume no anisotropic stress as discussed above.

Finally, in order to evaluate \refeq{gen4}, note that the Weyl tensor is trace-free over any two indices and has the same symmetry properties as the Riemann tensor.  Moreover, $\pd^F_k C_{Fl i j 0}= \mathcal{O}\left(\pd^F_k R_{Fl i j 0}\right)$ is a fluid-orthogonal derivative of the Riemann tensor and thus neglected as explained after \refeq{gen4}.  Moreover, in the CFC construction this term is naturally higher order, as for the leading expression given in \refeq{CFCleading}, $\pd^F_k C_{Fl i j 0}= \mathcal{O}\left(\vec{x}_F\right)$ vanishes when evaluated on the observer's worldline.  It follows then that the $0ij0-$component of the LHS of \refeq{gen4} becomes  
\begin{equation}\label{eq:weyllhs}
{-a_F}\left(\dot \E^F_{ij}  +H_F \E^F_{ij}\right)\,, 
\end{equation}
while all other components of \refeq{gen4} either vanish or are higher order
in derivatives, as shown in \refapp{Weyleqns}.  
Finally, for the RHS of \refeq{gen4}, we neglect spatial derivatives (in CFC) acting on the stress energy tensor and apply the continuity equation, \refeq{gen3F}, to yield on the observer's worldline
\be
4 \pi G a_F^2 \rho_F \sigma_{Fij}\,. 
\ee
Thus, without higher spatial derivatives and on the central geodesic,
\refeq{gen4} in CFC reduces to 
\begin{equation}
\dot{\E}_{Fij} +  H_F  \E_{Fij} + 4\pi G a_F  \rho_F \sigma_{Fij}=0\,.
\label{eq:gen4F}
\end{equation}
Note that \refeqs{weyllhs}{gen4F} would look different
if we had inserted the covariant definition 
$E_\mn (U) := C_{ \mu\alpha\nu \beta}U^\alpha U^\beta$ into \refeq{gen4}, as
the spatial derivatives acting on $U^\mu$ yield non-negligible terms.  
Since our goal is to derive a closed system in terms of \emph{local}
gravitational observables in the framework of the CFC, we choose 
\refeq{weyllhs} as the local approximation to \refeq{gen4}.  
In order to emphasize the subtle distinction between $E_\mn(U)$ and
$C^F_{i0j0}$ in the context of our local approximation, we have introduced 
the new symbol $\E^F_{ij}$ in \refeq{EFdef}.  

The physical interpretation of $\E^F_{ij}$ becomes clear when deriving 
its relation to the perturbation $h^F_{00}$ of the
00-component of the metric in CFC \refeq{CFCleading}.  Using
the transformation law of the Riemann tensor under a conformal
rescaling of the metric, one obtains
\begin{align}
\tilde{R}_{0l0m}^F =\:& \frac{1}{\af^2}{R^F}_{0l0m}+ \delta_{lm} \af^2  \left(\dot{H}_F+ H_F^2\right) \notag\\
=\:&   \frac{1}{\af^2}{R^F}_{0l0m}-\delta_{lm} \left(\frac{1}{3} \sigma_{Fij}\sigma_F^{ij} +\frac{4\pi G}{3}\af^2  \rho_F\right)\,,
\end{align}
where we have used \refeq{gen1F} in the second line.  Using the definition
of the Weyl tensor as trace-subtracted version of the Riemann tensor $\tilde R^F$,
we then obtain [see also Eq.~(3.31) in \cite{Dai:2015jaa}]
\begin{align}
h_{00}^F =\:& \left( - \E^F_{lm} + \frac{1}{3} \delta_{lm} \sigma_{Fij}\sigma_F^{ij}  \right)_{\v{0}} x_F^l x_F^m\,,
\label{eq:hF00}
\end{align}
where we have used \refeq{EFdef}.  Note that since the Weyl tensor is invariant under conformal rescaling, we have $\tilde C^F = C^F$.  We see that $\E^F_{lm} = C^F_{0l0m}/a_F^2$ is exactly
the trace-free part of $-\partial_l\partial_m h_{00}^F$, which is the
local tidal field acting on non-relativistic bodies in the CFC frame.  
Any other Ricci-contribution to the local tidal field would have to be due to anisotropic
stress.  


The four equations \refeqs{gen1F}{gen4F} now clearly form a closed, local system of ordinary differential equations governing the evolution of the four unknowns $\rho_F,\, H_F,\,\sigma_F,$ and $\E_F$ along the fluid geodesic:
\begin{align}
&\dot{H}_F+H_F^2 +\frac{1}{3 \af^2} \sigma_{Fij}\sigma_F^{ij} +\frac{4\pi G}{3} \rho_F=0
\label{eq:system}\\
&\dot{\sigma}_{Fij} + H_F \sigma_{Fij}+ \frac{1}{a_F}\left(\sigma_{Fil}\sigma_F^l{}_j -\frac{1}{3}\delta_{ij} \sigma_{Flm}\sigma_F^{lm}\right)+\frac{\E_{Fij}}{a_F}=0\vs
&\dot{\rho}_F  +3H_F \rho_F =0\vs
&\dot{\E}_{Fij} +  H_F  \E_{Fij} + 4\pi G a_F  \rho_F \sigma_{Fij}=0\,.
\nonumber
\end{align}
Note that $\E_F$ has dimensions 1/length$^2$, while $\sigma_F$ has dimensions 1/length.  
Given initial conditions for $\rho_F, H_F$ (or equivalently, curvature $K_F$), $\sigma^F_{ij}$, and $\E_{Fij}$, \refeq{system} can be integrated numerically without any further approximations.  Since we have used the Einstein equations
through \refeq{gen1} and \refeq{gen4} (with all components of the latter
derived in \refapp{Weyleqns}), all constraints are self-consistently 
fulfilled at leading order in derivatives.  
In the next section, we will consider the general perturbative solution around a flat matter-dominated (Einstein-de Sitter) universe.  

The fully relativistic system \refeq{system} clarifies the physical meaning of the locally observable impact of tidal fields.  That is, tides are a manifestation of the Weyl tensor, which encodes the nonlocal part of gravitational interactions, and thus of the large-scale \emph{inhomogeneities} in the matter distribution.  This is to be contrasted with \emph{homogeneous and anisotropic} Bianchi~I spacetimes, where the anisotropy is sourced by the trace-free part of the Ricci tensor $\hat R_{ij}$.  As we argued above, this in fact vanishes everywhere for a pressureless (and indeed any ideal) fluid.  Thus, the construction leading to \refeq{system}, rather than a Bianchi~I solution, are the proper relativistic model of long-wavelength density and tidal perturbations.

Note that the system simplifies further when setting $\sigma_F = 0 = \E_F$ initially, corresponding to a spherically symmetric system.  The symmetry is preserved so that $\sigma_F$ and $\E_F$ remain zero.  We then obtain
\begin{align}
& \dot{H}_F+H_F^2 +\frac{4\pi G}{3} \rho_F=0
\label{eq:systemsph}\\
&\dot{\rho}_F  +3H_F \rho_F =0\,.
\nonumber
\end{align}
These are just the second Friedmann equation and continuity equation of the FLRW spacetime.  One can then show (again, assuming scales much larger than the sound horizon of the fluid) that the first Friedmann equation is satisfied as well \cite{Dai:2015jaa}.  This proceeds in the usual way by multiplying \refeq{systemsph} with $\dot a_F$, integrating once, and using the continuity equation to yield
\begin{align}
H_F^2 = \frac{8\pi G}{3} \rho_F -\frac{K_F}{\af^2},
\end{align}
where $K_F$, the local spatial curvature, is an integration constant, hence is conserved.  
This is the well-known \emph{separate universe} picture: long-wavelength isotropic adiabatic perturbations are indistinguishable from an FLRW spacetime (with different cosmological parameters) by local observations.  \refeq{systemsph} is also equivalent to the spherical collapse equation \cite{sep1}.  Note that \refeq{systemsph} is exact given the said assumptions, while in the anisotropic case \refeq{system} is an approximation whose accuracy we will discuss further below.  First, however, we will proceed to solve the general, anisotropic case.

\section{Perturbative solution in Einstein-de Sitter}
\label{sec:sol}

We now consider a perturbed Einstein-de Sitter universe, where in the background the Hubble rate satisfies $\bar H^2 = 8\pi G\bar\rho/3  \propto \bar a^{-3}$.  
We expand all quantities into orders of perturbation, denoted by ${}^{[n]},n=1,2,\cdots$.  Quantities with an overbar are evaluated in the background Einstein-de Sitter universe.  Correspondingly, we write the density as
\be
\rho = \bar{\rho}(t)\left[1+ \delta_F\left(t,x^i\right)\right]\,.
\ee
Throughout, we assume that initial conditions are set at sufficiently early times (see \refsec{IC} below), so that we can restrict to the fastest growing modes throughout.  This is merely for calculational convenience;  it is straightforward to keep the subleading modes when solving \refeq{system}.  

\subsection{Initial conditions}
\label{sec:IC}

We briefly describe how the growing-mode initial conditions for 
\refeq{system} can be determined.  For this, we assume they are set
at sufficiently early times, so that linear perturbation theory is
accurate.  We can then relate the CFC-frame quantities to those
in a given global coordinate system at linear order in perturbations.  
Specifically, we consider two frequently used gauge choices:  
conformal-Newtonian gauge, defined through
\be
ds^2 = a^2(\tau)\left[- (1+2\Phi) d\tau^2
+ (1-2\Psi) \d_{ij} dx^i dx^j \right]\,,
\label{eq:metriccN}
\ee
and comoving gauge, which is frequently used for calculations during inflation:
\be
ds^2 = a^2(\tau)\left[- (1+2N_1) d\tau^2
+ 2 N_i d\tau dx^i
+ (1+2\mathcal{R}) \d_{ij} dx^i dx^j \right]\,.
\label{eq:metriccom}
\ee
Here, spatial slices are chosen such that $T^0_{\  i}=0$, hence the designation ``comoving.''  
In both cases, we have only included scalar perturbations.  The reason
is that vector perturbations are pure decaying modes in both cases,
so that they are irrelevant for the fastest growing modes.  Tensor modes
on the other hand are propagating modes whose nonlinear evolution we do not
expect to be described correctly by the local approximation employed in
\refeq{system}.  The linear evolution of a tensor-mode-induced $\E^F_{ij}$ 
and its effect on small-scale perturbations was derived in \cite{tidalpaper,CFCpaper}.  
We stress again that this simplification is merely for convenience and not required within 
the CFC formalism; one can straightforwardly include vector and tensor
modes and their associated decaying modes.  

Let us first consider the simpler,
isotropic case, where the initial conditions can be either specified
through $\delta_F$ or $K_F$.  As shown in \cite{Dai:2015jaa},
one has 
\be
K_F = \frac23 \left(\partial^2 \Psi - \cH \partial_i v^i \right) = \frac23 \partial^2 \mathcal{R}\,.
\ee
Note that $K_F = $~const at all times (not only during matter domination), as long as it is outside the sound horizon of all fluid components.  During
matter domination and at linear order, one further has
\be
K_F = \frac53 \Omega_{m0} H_0^2 a^{-1}(t) \delta_F(t)\,,
\ee
where $\Omega_{m0}$ is the matter density parameter today.

Next, consider the anisotropic case.  Restricting to the growing mode,
it is sufficient to provide initial conditions for $\E^F_{ij}$.  In
conformal-Newtonian gauge, we have at linear order in perturbations
\be
\E^F_{ij}\Big|_\text{linear} =  \left(\partial_i\partial_j - \frac13 \d_{ij} \partial^2\right)\Phi\,,
\label{eq:EFcN}
\ee
where we have used the CFC metric constructed at linear order by \cite{Dai:2015jaa}, and \refeq{hF00}.\footnote{Note that $g_{00}^F= a_F^2[-1 + h_{00}^F] = - a_F^2 [1-h_{00}^F]$.}  Moreover, using that the two spacetime potentials in conformal-Newtonian gauge are equal in the absence of anisotropic stress, and setting vorticity to zero, it is easy to show that \refeq{gen4F} yields the correct evolution of
$\E^F_{ij}$ \emph{at linear order} [cf. Eq.~(4.50) in \cite{Bertschinger:1993xt}].  

Note that unlike the curvature $K_F$, $\E^F_{ij}$ is in general not conserved during cosmic evolution even outside the sound horizon of all fluid components.  This is a qualitative difference to the isotropic case, where the effect of the 
long-wavelength perturbation is described by $K_F$ which is constant at all orders on large scales.  However, in the particular case of matter domination, $\E^F_{ij}$ is conserved at linear order (see below).  In this case, initial conditions can be simply specified by using the well-known relation for growing-mode adiabatic perturbations during matter domination, $\mathcal{R} = (5/3)\Phi$, so that
\be
\E^F_{ij}\Big|_\text{linear, mat. dom.} = \frac35 \left(\partial_i\partial_j - \frac13 \d_{ij} \partial^2\right) \mathcal{R}\,.
\label{eq:EFcom}
\ee
This relation can be used immediately to initialize a calculation for the nonlinear evolution of the tidal field $\E^F_{ij}$ for modes that enter the horizon during matter domination.  In general, one should follow the \emph{linear} evolution of $\E^F_{ij}$ via \refeq{EFcN} until matter domination (for example, using a Boltzmann code), at which point $\E^F_{ij}$ approaches a constant and can be matched to the \emph{nonlinear} calculation in matter domination which we describe next.  

\subsection{Perturbative solution up to second order}

We begin with the linear evolution in CFC, which matches that of standard
(subhorizon) perturbation theory \cite{bernardeau/etal}, and was
derived in \cite{Dai:2015jaa}.  In terms of the matter density perturbation
and electric Weyl tensor, we obtain
\begin{align}
\delta_F^\oo =\:& \bba(t_F)\dl(\vx_F) \\
\E_{Fij}^\oo =\:& \E_{Fij}^\oo(\vx_F) \,.
\end{align}
Here, we have introduced the linearly-extrapolated initial overdensity 
$\dl(\vx_F) := \d_F^{[1]}(\vx_F,t_0)$ and the local tidal field, $\E_{Fij}^\oo(\vx_F)$, for later
convenience.  Note that $\delta_F^\oo \propto \bba$, while $\E_{Fij}^\oo =$~const at linear order.  One could thus
think of $\E_{Fij}$ as the analog for anisotropic perturbations of the spatial
curvature $K$.  Unlike the latter however, we will see that $\E_{Fij}$ is
not conserved at nonlinear order.  Moreover, even at linear order the conservation of $\E_{Fij}$  only holds in a flat matter-dominated universe.  

The local scale factor and velocity shear are at linear order given by
\begin{align}
\af^\oo =\:& \bba(t_F) -\frac{1}{3} \bba^2(t_F) \dl(\vx_F) \\
\sigma_{Fij}^\oo =\:& - \bba^{1/2}(t_F)  t_0  \E_{Fij}^\oo(\vx_F) \,.
\end{align}
Note that $\bba(t_F)= (t_F/t_0)^{2/3}$-- the familiar scale factor in the unperturbed FLRW spacetime.  

Continuing to solve the equations \refeq{system} perturbatively, and keeping only the fastest growing mode (i.e. the term with the highest power of $t_F$), gives, at second order in perturbations,
\begin{align}
\af =\:& \bba-\frac{1}{3} \bba^2 \dl-\frac{1}{21} \bba^3\left(\dl\right)^2- \frac{3}{14}\bba^3 t_0^4\left(\E_{Fij}^\oo\right)^2+\cdots \label{eq:aF}\\
\delta_F =\:& \bba\dl+\frac{17}{21}\bba^2{\dl}^2+\frac{9}{14} \bba^2 t_0^4\left(\E_{Fij}^\oo\right)^2+\cdots \label{eq:deltaF}\\
\E_{Fij} =\:& \E_{Fij}^\oo + \frac{20}{21}\bba\dl \E_{Fij}^\oo + \frac{3}{7}\bba    t_0^2\left(\E_{Fil}^\oo {\E_F^l{}_j}^\oo -\frac{1}{3}\delta_{ij} \E_{Flm}^\oo{\E_F^{lm}}^\oo\right) +\cdots \\
\frac{\sigma_{Fij}}{\bba^{1/2}} =\:& - t_0 \E_{Fij}^\oo - \frac{19 }{21} \bba \dl t_0 \E_{Fil}^\oo\vs
&  -\frac{18 }{21}  \bba\, t_0^3  \left(\E_{Fil}^\oo {\E_F^l{}_j}^\oo -\frac{1}{3}\delta_{ij} \E_{Flm}^\oo{\E_F^{lm}}^\oo\right)    +\cdots\,,
\label{eq:sigmaF}
\end{align}

where the left-hand side is evaluated at $t_F$ and $x_F$?, while on the right-hand side the time dependence is completely encoded in the factors of $\bba =\bba(t_F)$.  
Instead of writing the solutions in terms of the initial conditions at some
early time, we have followed the customary choice in cosmological perturbation
theory of phrasing the solution in terms of the linearly-extrapolated initial overdensity $\dl(\vx_F)$ and initial electric Weyl contribution $\E_F^{[1]}$ introduced above.    
The results \refeqs{aF}{sigmaF} can be formally continued straightforwardly to higher
order.  Further, in order to describe the solution in a $\Lambda$CDM background, one can perform the standard very accurate approximation of replacing
$\bba^n$ with the linear growth factor $[D(t_F)]^n$.

\subsection{Connection to standard perturbation theory}
\label{sec:SPT}

It is instructive to compare our result \refeq{deltaF} to the second order density perturbation in standard (subhorizon) perturbation theory \cite{bernardeau/etal,sherwin/zaldarriaga}, which yields
\be \label{eq:sherdel}
\d(\vx,t) = \d^{[1]}(\v{q}[\vx,t],t) + \frac{17}{21} (\d^{[1]})^2 
+ \frac27 (K_{ij}^{[1]})^2\,,
\ee
where $\v{q}[\vx,t]$ denotes the Lagrangian position corresponding to the given Eulerian coordinate, and 
\be
K_{ij}^{[1]} := \frac1{4\pi G\rhob} \left[ \partial_i \partial_j - \frac13 \partial^2\right]\Phi
=  \frac{2}{3}\frac{1}{\bba^2 \bar{H}^2} \E_{Fij}^\oo 
=  \frac32 \bba t_0^2 \E_{Fij}^\oo 
\,,
\label{eq:Kijdef}
\ee 
where we have used the Einstein-de Sitter background.  
The second term in \refeq{sherdel} describes the second order growth of density perturbations in the absence of tidal fields; its coefficient is exactly what is obtained from the second order expansion of the spherical collapse solution.  The third term encodes the tidal effects on the density perturbations.  In standard Eulerian perturbation theory, the first, linear term is expanded around the Eulerian position yielding a displacement term
\be
\d^{[1]}(\v{q}[\vx,t],t) = \d^{[1]}(\vx,t) - \v{s}^{[1]}(\vx,t) \cdot\boldsymbol{\nabla}  \d^{[1]}(\vx,t)\,,
\ee
where $\v{s}^{[1]}$ is the linear displacement from the Lagrangian to the Eulerian position.  Since the CFC calculation corresponds to working in Lagrangian coordinates (as the origin of the coordinate system comoves with the fluid, evidenced by the fact that $\v{v}_F=0$ on the geodesic), this displacement does not appear in \refeq{deltaF};  that is, the CFC formalism automatically resums all the displacement terms appearing in Eulerian perturbation theory.  
Using that 
\be
\frac{9t_0^4}{14} \bba^2 \left(\E_{Fij}^\oo\right)^2
= \frac9{14} \frac49 (K_{ij}^\oo)^2 = \frac27 (K_{ij}^\oo)^2\,,
\ee
we immediately see that \refeq{deltaF} matches the standard perturbative
calculation \refeq{sherdel}.  However, in the derivation of \refeq{deltaF}
we have not assumed that the scale of the perturbation is much smaller than
the Hubble horizon.  This shows that, when measured in the
proper rest frame, the second order evolution of the matter density
in the presence of tidal fields is \emph{exactly} as given by the standard
result which is seemingly only valid on subhorizon scales.  This fact
was already known for isotropic perturbations, in which case the evolution
is determined by a local Friedmann equation \cite{Dai:2015jaa}.  \refeq{deltaF} generalizes
this result to the anisotropic case, i.e. the inclusion of tidal fields.  

However, the agreement between the evolution predicted by the closed
system \refeq{system} and standard perturbation theory no longer
holds at higher order.  This can already be seen in the result for
$\sigma_{Fij}$ at second order, \refeq{sigmaF}.  The SPT prediction
for $\sigma_{ij}$, which corresponds to the derivative with respect to
Lagrangian coordinates of the fluid velocity $\v{v}$, can be derived
by using that $\v{v}$ is equal to the convective (or Lagrangian)
time derivative w.r.t. $\tau$ of the Lagrangian displacement $\v{s}$.  This yields
(e.g. App.~B of \cite{MSZ}) 
\ba
\frac{\sigma_{ij}}{\bba^{1/2}} =\:& - t_0 \E_{Fij}^\oo 
+ \bba^{1/2} \left[\frac{\partial_i\partial_j}{\nabla^2} - \frac13 \d_{ij} \right]
 \frac{\partial}{\partial\tau} \left[-\frac2{14} (\d^{[1]})^2 + \frac3{14} (K^{[1]}_{ij})^2 \right] \vs
 =\:& - t_0 \E_{Fij}^\oo 
+ \bba \left[\frac{\partial_i\partial_j}{\nabla^2} - \frac13 \d_{ij} \right]
\left[-\frac4{21} t_0^{-1} (\dl)^2 
+ \frac9{14} t_0^3 \, \E_{Flm}^\oo{\E_F^{lm}}^\oo \right] \,.
\label{eq:sigmaSPT}
\ea
Clearly, this differs from \refeq{sigmaF}.  In particular, the SPT result
is \emph{spatially nonlocal} (the same holds when deriving the SPT result
for the nonlinear evolution of $\E_{Fij}$).  The differences go back to the terms
neglected when evaluating the evolution equation \refeq{gen4} for the electric
Weyl tensor component.  When neglecting these terms, we were able to obtain
a closed system that is local in space around the fluid geodesic.  However,
the gravitational evolution 
of density and tidal fields within a region, when followed over a finite 
duration of time, is not completely described by the local tidal and density
field.  This is encoded in the nonlocal terms appearing at second order in
$\sigma_{ij}$ in standard perturbation theory, which, apart from dropping
post-Newtonian terms, does not make approximations in \refeq{gen4}.  Note that 
the nonlocal term appears in the density
perturbation $\d_F$ only at third order (see also \cite{Noh:2003yg, Hwang:2015jja});  in fact the nonlocal contributions
to $\d_F^{[3]}$ are proportional to $\sigma^{[1] lm} \sigma_{lm}^{[2]}$ \cite{MSZ}.  Finally, we see that the nonlocal terms disappear in the case of 
spherical symmetry, in which case \refeq{system} recovers the exact
separate universe result which matches perturbation theory to all orders.  

In the next section, we will connect our results to those from other,
previously considered local approaches to nonlinear gravitational evolution
of non-spherically symmetric systems.

\subsection{Connection to the local tidal approximation and ellipsoidal collapse}
\label{sec:LTA}

Interestingly, the system \refeq{system} has been derived before,
in the subhorizon limit, by Ref.~\cite{Hui:1995bw} who referred to this
as the \emph{Local Tidal Approximation} (LTA).  This approximation
originated in a series of attempts at improving the local approximation to study large-scale structure and at clarifying  the relation between general relativity and Newtonian dynamics.  The first was the  Zel’dovich approximation (ZA) \cite{1970A&A.....5...84Z}.  
Bertschinger and Jain \cite{1994ApJ...431..486B} proposed the
non-magnetic approximation (NMA).  
In the NMA, the covariant magnetic part of the Weyl tensor is set to zero, $H_{ij} =0$.  However, Refs.~\cite{1994ApJ...435....1B,1995ApJ...442...30K} showed that $H_{ij}$ cannot be consistently neglected in the Newtonian limit.  That is, $H_{ij}$ has a Newtonian counterpart after all. 
Later, Hui and Bertschinger proposed the LTA. Here they defined a new quantity $M_{ij}$, composed of a combination of terms in the tidal evolution equation,  and set it to $0$.
They found that whereas NMA is exact for spherical perturbations but not cylindrical ones; LTA is exact for both and, more generally, for any growing-mode perturbations whose gravitational equipotential surfaces have constant shape with time. 

In the LTA approximation, the authors imposed the condition 
\begin{equation}
M_{ij} := -\nabla_k \epsilon^{kl}{}_{(i}H_{j)l} +\theta E_{ij} +\delta_{ij} \sigma^{kl} E_{kl} -3 \sigma^k{}_{(i}E_{j)k} -\omega^k{}_{(i}E_{j)k} = 0\,,
\label{eq:LTA1}
\end{equation}
where conformal-Newtonian gauge is adopted and $E_{ij} \equiv (\partial_i\partial_j - (\d_{ij}/3)\nabla^2)\Phi$ is the tidal field in the subhorizon approximation.  Note that in the subhorizon limit, the velocity-orthogonal projection is equivalent to simply restricting to spatial coordinates.  
$H_{ij}$ here is the magnetic component of the Weyl tensor. The tensor $M_{ij}$ can also be written as
\begin{equation}
M_{ij}  = -4\pi G a^2 \rho \nabla_{(i}v_{j)}- \frac{\diff }{\diff t} \left(\nabla_i\nabla_j a \frac{h_{00}}{2} \right).
\label{eq:LTA2}
\end{equation}
Using \refeq{hF00} for $h_{00}$ in CFC, \refeq{gen4F} implies immediately
that the trace-free part of $M_{ij}$ vanishes.  On the other hand, the trace
part corresponds to a combination of the Raychaudhuri and continuity equations. 
Correspondingly, \refeq{system} agrees with Eq.~(22) of \cite{Hui:1995bw}.  
Thus, the CFC approach offers a simple and fully covariant derivation of the LTA.  In addition, it yields an additional interpretation of the nature of the
approximation made in the LTA construction, namely through the terms neglected in \refeq{gen4}. The resultant $E_{ij}$  in LTA is in fact our local tidal field $\E^F_{ij}$.

Ref.~\cite{Hui:1995bw} discussed the LTA in the context of the collapse
of a homogeneous \emph{isolated} ellipsoid \cite{icke:1973,white/silk:1979}.  
In this model, one neglects the gravitational effect of the ellipsoid on
the surrounding matter.  Ref.~\cite{White:1994bn} studied the validity
of this approximation by performing an N-body simulation which allows for
the backreaction of the ellipsoid on its environment.  Specifically, the environment consisted of a small homogeneous negative density perturbation to compensate for the mass contained within the overdense ellipsoid.  
He found that the above-mentioned approximation is
quite accurate until the late stages of collapse.  The advantage of neglecting
gravitational backreaction on the environment is that a closed, semi-analytical solution can be obtained.  

In particular,
working in the conformal-Newtonian gauge and taking the subhorizon limit, the evolution of an irrotational, isolated homogeneous ellipsoid embedded in an FLRW background is governed by 
\begin{align}
\frac{\diff^2 R_i}{\diff t^2} = -2\pi G R_i \left[\frac{2}{3} \bar{\rho} +\alpha_i \delta\right], \quad\tn{for }i=1,2,3,
\label{eq:Eeom}
\end{align}
where $R_i$ are the proper axis lengths of the ellipsoid; $\bar{\rho}$ is the local homogeneous (or, mean) density around the ellipsoid,  such that $\rho := \bar{\rho}\left(1+\delta\right)$ ; $t$ is the proper time; and $\alpha_i$ are defined by
\begin{align}
\alpha_i :=  \left(\prod_{n=1}^3 R_n \right) \int\limits_{0}^{\infty} \frac{\diff s}{\left(R_i^2 +s\right)\sqrt{\sum\limits_{k=1}^3 \left(R_k^2+s\right)}}\,.
\label{eq:alphaeom}
\end{align}
The peculiar velocity field inside the ellipsoid is 
\begin{align}
v_i = a\left(\frac{\dot{R}_i}{R_i}-\frac{\dot{a}}{a}\right)x_i,
\end{align}
where $x_i$ are comoving spatial coordinates.  
These relations can be used to derive the local Hubble rate, velocity shear, and
electric Weyl tensor component \cite{Hui:1995bw}, which in CFC become
\begin{align}
&H_F=\frac{1}{3}\sum\limits_i \frac{\dot{R}_i}{R_i}\\
&\sigma_{Fij} = a_F\diag\left(\frac{\dot{R}_i}{R_i}- H_F \right)\\
&E_{Fij} = 2\pi G a_F^2 \delta_F\,\, \diag\left(\alpha_i -\frac{2}{3}\right)\,.
\end{align}
Given this one-to-one mapping to quantities in the system \refeq{system}, 
the isolated ellipsoid can be considered as another local approximation 
to the evolution of non-spherically symmetric perturbations, even though
it is not truly local as the $\alpha_i$, and thus $\E_{Fij}$, obey an
integral equation \refeq{alphaeom}.  
As shown in \cite{Hui:1995bw} however, the LTA, and hence the closed system
in CFC derived in \refsec{eqns:CFC}, recovers the evolution of 
\refeqs{Eeom}{alphaeom} only up to first order.  
Thus, given the N-body results of \cite{White:1994bn}, the LTA describes the
actual evolution of an isolated ellipsoidal perturbation much less 
accurately than \refeqs{Eeom}{alphaeom}.  This difference can be 
attributed to the perfectly local approximation made in LTA, whereas
the isolated ellipsoid does take into account the finite extent of the 
perturbation.  

However, in practice, perturbations in the universe cannot be considered
isolated, at least until they have reached sufficient density to 
dominate the local gravitational potential.  
As discussed in \refsec{SPT}, the correct evolution of tidal fields
is fundamentally nonlocal starting at second order.  This is qualitatively different from isotropic perturbations, where the LTA reduces to the ``separate universe'' which provides an exact solution for large-scale adiabatic perturbations.

\section{Application: the rest-frame matter three-point function}
\label{sec:bispectrum}

As derived in \refsec{sol}, the matter density perturbation in CFC frame
up to second order in perturbations is given by
\be
\d_F(\v{q},t) = D(t) \d_L^{[1]}(\v{q})
+ \frac{17}{21} [D(t) \d_L^{[1]}(\v{q})]^2 + \frac27 [D(t) K_{ij}^{[1]}(\v{q})]^2\,,
\ee
where $K_{ij}^{[1]}$ is defined through \refeq{Kijdef}. 
Here, we have only assumed adiabatic Gaussian initial conditions.  $\v{q}$
is a spatial coordinate which labels the geodesic around which the CFC
frame is constructed, while $t$ denotes proper time along the geodesic.
We will choose $\v{q}$ to denote the spatial position on a constant-proper-time
slice at $t=0$, that is, at the end of inflation.

We now would like to relate this result to the observed, late-time statistics
of the matter density field, without resorting to the sub-horizon approximation
made in most large-scale structure studies;  specifically, we are interested
in the leading signature of nonlinear evolution, the three-point function. 
Unfortunately, the precise
prediction for these statistics depends on how the matter density field
is measured, for example through gravitational lensing (which strictly
measures the deflection of photon geodesics), or tracers such as galaxies
(whose relation to matter is complicated by bias, redshift-space
distortions, and other effects; see \cite{biasreview} for a review). 
Since these issues go beyond the scope of this paper, we here assume
the idealized case of observers on different geodesics communicating
their local rest-frame matter density as well as proper time to a
distant observer on their future light cone.  

Let us define the displacement $\v{s}$ as parametrizing the difference
between the spatial position $\vx$, as observed by the distant observer, of the fluid geodesic relative to the initial position $\v{q}$ as a function of proper time:
\be
\v{x} = \v{q} + \v{s}(\v{q},t)\,.
\ee
Note that $\v{s}$ depends on how the distant observer measures the spacetime
position of the geodesic. In general, $\v{s}$ describes the effect of
large-scale gravitational perturbations on the geodesic, which are given by
an integral over the gradient of the gravitational potential, as well as the
details of the distant observer's measurement.  
For example, if they use apparent photon arrival
directions and redshifts, then $\v{s}$ includes Doppler shift, gravitational
redshift, and deflection by structures along the line of sight.  The explicit
expression of $\v{s}$ further depends on the coordinate (gauge) choice that
is used to calculate the displacement, although the end result is independent of the gauge; see \cite{jeong/schmidt:2015} for
a brief overview.  For this reason, we will not derive $\v{s}$ explicitly
here.  However, independently of the gauge choice, $\v{s}$ is
first order in cosmological perturbations.  Hence, we can expand
the CFC-frame density field up to second order as
\be
\d_F(\v{x},t) = \d_F(\v{q},t) - s^i(\v{q},t) \partial_i \d_F(\v{q},t)\,.
\ee
Moreover, since only the cross-correlation of $\v{s}$ with the density
field $\d_F$ enters in the leading three-point function, it is sufficient
to consider the longitudinal contribution to $\v{s}$, which we write
as
\be
s^i(\v{q},t) = \partial^i S(\v{q},t)\,,
\ee
where $S$ is a scalar function which, for a fixed Lagrangian position $\v{q}$, only depends on proper time $t$.  Note that $S$ has dimension of length$^2$.  

With a slight generalization of the results in App.~A of \cite{bel/hoffmann/gaztanaga}, we can then write the three-point function of the CFC-frame matter density as
\ba
\< \d_F(\v{x}_a,t)\d_F(\v{x}_b,t)\d_F(\v{x}_c,t) \> =
\biggl\{\: &
\frac{34}{21}\xi_0(r_{1}, t)\xi_0(r_{2}, t)
+ \frac47 \left[\mu_{12}^2 - \frac13 \right] \xi_2(r_{1}, t) \xi_2(r_{2}, t)
\vs
&
-
\mu_{12}
\left[ \frac{\partial\xi_{S\delta}(r_{1}, t)}{\partial r_1} \frac{\partial\xi_0(r_{2}, t)}{\partial r_2}
+
(1\leftrightarrow2)
\right]
\biggl\} \vs
+ &\: 2~\mbox{cycl. perm.}\,.
\label{eq:xim3}
\ea
Here, $\v{r}_1 = \v{x}_b-\v{x}_a$, $\v{r}_2 = \v{x}_a-\v{x}_c$, $\v{r}_1 = \v{x}_c-\v{x}_b$, while $\mu_{ij} \equiv \v{r}_i\cdot\v{r}_j/r_ir_j$, and 
we have defined
\ba
\xi_0(r, t) =\:& \int \frac{d^3\v{k}}{(2\pi)^3} P_{\rm L}(k, t)j_0(kr) \vs
\xi_2(r, t) =\:& \int \frac{d^3\v{k}}{(2\pi)^3} P_{\rm L}(k, t)j_2(kr) \vs
\xi_{S\delta}(r, t) =\:& \int \frac{d^3\v{k}}{(2\pi)^3} P_{S \delta}(k, t)j_0(kr)\,.
\ea
Finally, $P_{\rm L}(k, t)$ is the power spectrum of the linearly extrapolated
rest-frame matter density perturbations (this is equivalent to the
density perturbations in synchronous-comoving gauge), while $P_{S\delta}(k, t)$
is the cross-power spectrum between $S$ and $\delta$, again at linear
order.

While \refeq{xim3} only represents an idealized case, which does not
include the mapping from the local rest-frame matter density to a
realistic observable, it clearly illustrates the simplicity of the result
obtained, without restriction to subhorizon scales, when using
CFC to describe nonlinear gravitational evolution.  Given the
assumption stated at the beginning of the section, \refeq{xim3} is
fully accurate at second order, and not restricted to the squeezed limit.

\section{Summary and discussion}
\label{sec:concl}

The conformal Fermi coordinates (CFC) are a convenient construction designed
to explicitly isolate the leading locally observable gravitational effects
of long-wavelength perturbations in the cosmological context.  We show that using this construction,
there is a natural way to derive a closed system of ordinary differential
equations [\refeq{system}] describing the evolution of a long-wavelength perturbation.  Here,
long-wavelength means that the scale of the perturbation is much larger than
the sound horizon of all fluid components (note that the sound horizon for
pressureless matter is the nonlinear scale, $r_s \sim 1/k_{\rm NL}$ \cite{Baumann}).  This system is exactly equivalent to the local tidal approximation (LTA).  
Moreover, the CFC frame provides a fully relativistic realization of 
this closed system;  that is, the results are valid on horizon or super-horizon scales without any post-Newtonian corrections.  As shown in \cite{CFCorig,Dai:2015jaa}, the CFC also allows for a direct connection to the initial conditions 
from inflation at nonlinear order.

This construction clarifies the anisotropic generalization of the
separate universe result for spherically symmetric long-wavelength perturbations.  That is, while the latter are entirely described locally by a curved FLRW spacetime, anisotropic long-wavelength perturbations contain a nonzero electric
component $E_{\mn}$ of the Weyl tensor as well.  Physically, this is entirely different from a Bianchi~I spacetime, where the anisotropy is encoded in the tracefree part of the spatial \emph{Ricci} tensor.  The latter is negligible for ideal fluids.  We thus argue that the Bianchi~I picture is not the proper local physical representation of long-wavelength perturbations in our Universe.  

Solving the system of evolutionary equations up to and including second order, we have found that the solution
for the density field is exactly equivalent to the result of standard,
sub-horizon perturbation theory.  Since our closed system in the CFC frame 
recovers the exact nonlinear evolution of \emph{isotropic} perturbations,
and recovers the correct \emph{linear} evolution of anisotropic perturbations,
our result for $\delta_F$ represents the proper rest-frame matter density
at second order in perturbations, including all relativistic corrections (assuming matter
domination holds, and that primordial decaying modes can be neglected).  
This follows from the fact that anisotropic perturbations only contribute
to the density at quadratic or higher order.  Note that this statement includes any vector and tensor metric perturbations
that are generated by anisotropic perturbations at second order \cite{Matarrese:1997ay,baumann/etal:2007,Boubekeur:2008kn};  these
can only contribute to the rest-frame matter density at third order (we did
not however include \emph{primordial} vector and tensor modes, which would contribute
at second order).   A corollary is that, if we interpret
the results of standard N-body simulations in this frame, any post-Newtonian 
corrections to the measured density field have to be third order in
perturbation theory.  This provides a strong constraint on their numerical
relevance.  On the other hand, our closed system fails for local tidal
fields and velocity shear already at second order.  Further, 
post-Newtonian corrections to non-ideal fluids,
such as neutrinos, and on gravitational lensing are in general less suppressed
than those for the matter density \cite{bruni/thomas/wands,adamek/etal:2013,Adamek:2016zes}.  

In this context, one might wonder whether this approach 
could be used to study the generation of gravitational waves from
large-scale structure.  
However, for this one needs to define what gravitational wave means.  
A natural, physical definition is to derive the transverse-traceless 
component of metric perturbations in the far-field limit \cite{carbone/matarrese,carbone/etal}.  
Unfortunately, the local nature of the CFC construction implies that
we cannot derive the far-field limit of these metric perturbations in this
approach. 

As a first, simple application of these results, we have derived the leading three-point function of the CFC-frame matter density field in \refsec{bispectrum} [\refeq{xim3}]. While this expression is idealized in the sense that it assumes that local observers directly communicate their local density to a distant observer, it is fully valid on arbitrarily large scales, and not restricted to specific configurations such as the squeezed limit. 
A further interesting possible application of these results is the implementation
of N-body simulations with a long-wavelength, non-spherically symmetric 
perturbation imposed;  that is, the anisotropic generalization of 
the ``separate universe simulations'' presented in 
\cite{mcdonald:2003,sirko:2005,gnedin/kravtsov/rudd:2011,li/hu/takada:2014,wagner/etal:2014,baldauf/etal:2015}.  In principle, the effect of the electric
Weyl tensor component $\E_{Fij}$ can be simply included by adding an 
external force $\propto \E_{Fij} x^j$ to any particle with position $\vx$.  
This however breaks the periodic boundary conditions inherent in 
conventional N-body simulations, and can thus cause numerical problems.  
We leave this issue for future work.  Note also that, unlike the
isotropic case, where the superimposed long mode can be made nonlinear \cite{wagner/etal:2014}, this is nontrivial in the anisotropic case, as
the correct nonlinear evolution of $\E_{Fij}$ is nonlocal [see discussion 
after \refeq{sigmaSPT}].  

This is indeed the main caveat to all local approximations to the nonlinear
evolution of (non-spherically symmetric) perturbations.  At second
order, the gravitational evolution of tidal field and velocity shear is 
spatially nonlocal.  This comes as no surprise, given that we have
shown that the tidal forces are due to the Weyl tensor, which captures
those parts of the full Riemann tensor that are not locally related to 
the matter distribution.  In case of the density field, this nonlocality
only appears at third order.  Thus, in all applications restricted to
sufficiently low order, local approximations are exact;  however, 
starting at third order in the matter (and galaxy) density field,
spatial nonlocality is an unavoidable feature of nonlinear gravitational
evolution.

\acknowledgments
We would like to thank Marco Bruni, Giovanni Cabass, Liang Dai, 
Eiichiro Komatsu, Sabino Matarrese, Mehrdad Mirbabayi, 
and Simon White for enlightening discussions. SI would further like to thank Simon Nakach for helpful discussions.  FS acknowledges support from the Marie Curie Career Integration Grant  (FP7-PEOPLE-2013-CIG) ``FundPhysicsAndLSS.''

\appendix

\section{Completeness of \refeq{system}}
\label{app:Weyleqns}

We now show that the, perhaps surprisingly, simple form of the evolution equation for $\E^F_{ij}$ is the single nontrivial component of \refeq{gen4} in
CFC at leading order in derivatives.  This implies that \refeq{system}
consistently contains all constraints from the \emph{full} covariant Einstein system, with no additional constraints at leading order in derivatives.  
We begin with the full covariant equation for the Weyl tensor, \refeq{gen4} :
\begin{equation}
\nabla^\kappa C_{\mn \kappa\lambda}= 8\pi G \left(\nabla_{[\mu }T_{\nu ]\lambda}+\frac{1}{3}g_{\lambda[\mu }\nabla_{\nu ]}g_{\alpha\beta}T^{\alpha\beta}\right)\,,
\end{equation} 
evaluated in CFC, and consider all the distinct combinations of space-time indices, up to symmetries of the Weyl tensor (which are the same as those of the Riemann tensor). Henceforth, we shall drop the "F" subscript. CFC shall be assumed throughout.  The line element of the CFC metric is given by \refeq{CFCmetric}
\begin{equation}\label{eq:met}
g_\mn (x^\mu ) = a^2 (\tau) \left[\eta_\mn +h_\mn \left(\tau,x^i \right)\right],
\end{equation}
We shall first consider the LHS of the system,
\begin{align}
\nabla^\kappa C_{\mn \kappa\lambda} =& \nabla_\kappa C_\mn{}^ \kappa{}_\lambda \\
=& \pd_\kappa C_\mn{}^ \kappa{}_\lambda - \Gamma^\alpha_{\kappa\mu} C_{\alpha\nu}{}^ \kappa{}_\lambda
- \Gamma^\alpha_{\kappa\nu} C_{\mu\alpha}{}^ \kappa{}_\lambda
+\Gamma^\kappa_{\kappa\alpha} C_{\mu\nu}{}^\alpha{}_\lambda
- \Gamma^\alpha_{\kappa\lambda} C_{\mu\nu}{}^ \kappa{}_\alpha.
\end{align}
The relevant Christoffel symbols of \refeq{met}  evaluated on the geodesic are \cite{CFCpaper},
\begin{align}
\Gamma^\mu_{0\nu}\rvert_\tn{geo} = \mathcal{H}\rvert_\tn{geo} \delta^\mu_\nu , \quad
\Gamma^0_{ij}\rvert_\tn{geo} = \mathcal{H}\rvert_\tn{geo} \delta_{ij}
\end{align}
For $\{\mu,\nu,\lambda\} = \{0,0,0\}$, $ C_{ 00 \kappa 0}$ is automatically zero given the symmetries of the Weyl tensor. 
For  $\{\mu,\nu,\lambda\} = \{0,0,i\}$, the only non-vanishing Weyl component is $ C_{0i k 0}$. Since $h_\mn =\mathcal{O}(x^2)$ and in leading order CFC we restrict to 2 spatial derivatives, regarding spatial derivatives acting on the metric, only terms of the form $\pd_k^2 h_\mn$ survive. Also, any spatial derivative of the Weyl tensor is neglected in leading order CFC. We further note that the Weyl tensor is tracefree w.r.t. any 2 indices. It follows that 
\begin{align}
\nabla^\kappa C_{ 0i \kappa 0} =0.
\end{align}
For  $\{\mu,\nu,\lambda\} = \{0,i,j\}$, we retrieve the LHS of the evolution equation for $\E_{ij}$ in \refeq{system}.
For  $\{\mu,\nu,\lambda\} = \{i,j,k\}$, we have
\begin{align}
\pd_\tau C_{ij}{}^0{}_k\,.
\end{align}
We shall now consider the RHS of \refeq{gen4} for the components with non-trivial LHS, adopting an ideal pressureless fluid stress-energy tensor (again, we really only require the absence of pressure perturbations).  
For  $\{\mu,\nu,\lambda\} = \{0,i,j\}$, we retrieve the RHS of the evolution equation for $\E_{ij}$ in \refeq{system}.
For $\{\mu,\nu,\lambda\} = \{i,j,k\}$, noting that $U_i\rvert_\tn{geo}=0$ (when not acted on by a spatial derivative),
\begin{align}
8\pi G \left(\nabla_{[i }T_{j ] k}+\frac{1}{3}g_{k [i }\nabla_{j ]}g_{\alpha\beta}T^{\alpha\beta}\right)
=&-\frac{ 8\pi G a^2}{3}\delta_{k [i }\pd_{j ]}\rho = 0 \quad
\mbox{(higher derivative)}\,.
\end{align}
This is a higher-derivative contribution, as it involves a spatial 
(fluid-orthogonal) derivative on the stress-energy tensor and is thus
equivalent to a third spatial derivative acting on the metric.  
Consequently, the only apparently non-trivial component of the system, besides our equation for $\E_{ij}$, is
\begin{align}
\dot{C}_{ij}{}^0{}_k =0\,.
\end{align}
At this order in derivatives, this component is entirely decoupled from the other quantities in \refeq{system} and corresponds to a constant that can be set to zero.

\bibliographystyle{jhep}
\bibliography{ref}

\end{document}